\documentclass[10pt,conference,letterpaper]{IEEEtran}
\usepackage{graphicx}
\usepackage{latexsym}
\usepackage{tabularx}
\usepackage{cite}
\usepackage{algorithm}
\usepackage{color}
\usepackage{amsfonts,amssymb}
\usepackage{amsthm,amsmath}
\usepackage{url}
\usepackage[encapsulated]{CJK} 
\usepackage{indentfirst}
\usepackage{soul}
\usepackage{subfigure}

\usepackage[paper=letterpaper,left=0.64in,right=0.64in,top=0.75in,bottom=1.05in]{geometry}

\usepackage{enumitem}
\setlist[enumerate]{leftmargin= 0.5 cm}
\setlist[itemize]{leftmargin=0.3 cm}

\usepackage{algorithmicx}
\usepackage{algorithm,algpseudocode}

\makeatletter
\newcommand\subparagraph{%
  \@startsection{subparagraph}{5}
  {\parindent}
  {3.25ex \@plus 1ex \@minus .2ex}
  {-1em}
  {\normalfont\normalsize\bfseries}}
\makeatother

\makeatletter
\algrenewcommand\algorithmicindent{0.2 cm}

\algnewcommand\algorithmicinput{\textbf{Input:}}
\algnewcommand\INPUT{\item[\algorithmicinput]}

\algnewcommand\algorithmicoutput{\textbf{Output:}}
\algnewcommand\OUTPUT{\item[\algorithmicoutput]}

    \newcounter{phase}[algorithm]
    \newlength{\phaserulewidth}
    \newcommand{\setphaserulewidth}{\setlength{\phaserulewidth}}
    
    \setphaserulewidth{.7pt}
\makeatother

\theoremstyle{definition}

\theoremstyle{plain}
\newtheorem{theo}{Theorem}

\newtheorem{coro}{Corollary}

\graphicspath{ {images/} }

\IEEEoverridecommandlockouts
\begin{document}
\title{Scalable Rate Control for Traffic Engineering with Aggregated Flows in Software Defined Networks\vspace{-0.1 cm}}
\author{
Jian-Jhih Kuo\IEEEauthorrefmark{1}, Chih-Hang Wang\IEEEauthorrefmark{2}, Cheng-Da Tsai\IEEEauthorrefmark{2}, De-Nian Yang\IEEEauthorrefmark{1}, and Wen-Tsuen Chen\IEEEauthorrefmark{1}\IEEEauthorrefmark{2}\\
\IEEEauthorrefmark{1}Institute of Information Science, Academia Sinica, Nankang, Taipei, Taiwan\\
\IEEEauthorrefmark{2}Department of Computer Science, National Tsing Hua University, Hsinchu, Taiwan
\thanks{\IEEEauthorrefmark{1} E-mail: \{lajacky,dnyang,chenwt\}@iis.sinica.edu.tw}
\thanks{\IEEEauthorrefmark{2} E-mail: s100062591@m100.nthu.edu.tw, s103062629@m103.nthu.edu.tw}\vspace{-0.3cm}}

\maketitle

\begin{abstract}
To increase the scalability of software defined networks (SDNs), flow aggregation schemes have been proposed to merge multiple mouse flows into an elephant aggregated flow for traffic engineering. In this paper, we first notice that the user bit-rate requirements of mouse flows are no longer guaranteed in the aggregated flow since the flow rate decided by TCP fair allocation is usually different from the desired bit-rate of each user. To address the above issue, we present a novel architecture, named Flexible Flow And Rate Management (F$^2$ARM), to control the rates of only a few flows in order to increase the scalability of SDN, while leaving the uncontrolled flows managed by TCP. We formulate a new optimization problem, named Scalable Per-Flow Rate Control for SDN (SPFRCS), which aims to find a minimum subset of flows as controlled flows but ensure that the flow rates of all uncontrolled flows can still satisfy minimum required rates by TCP fair allocation. We prove that SPFRCS is NP-hard and design an efficient algorithm, named Joint Flow Selection and Rate Determination (JFSRD).  Simulation results based on real networks manifest that JFSRD performs nearly optimally in small-scale networks, and the number of controlled flows can be effectively reduced by 50\% in real networks.
\end{abstract}

%\begin{IEEEkeywords}
%software defined network (SDN), traffic engineering, flow aggregation, rate control, tcp, np-hard 
%\end{IEEEkeywords}

\IEEEpeerreviewmaketitle
%\vspace{-0.2 cm}
\section{Introduction}
Software Defined Network (SDN) is an emerging network paradigm that separates the control plane from the data plane through OpenFlow \cite{OpenFlow}. This decoupling abstracts lower level functions into higher level services and allows the policy enforcement and network configuration to be very flexible. Moreover, SDN provides a global view of the network and thus enables optimal decisions in a centralized manner to meet the diverse traffic demands from various application services. 

In this paper, we first notice that the user bit-rate requirements of mouse flows are no longer guaranteed in the aggregated flow since the flow rate decided by TCP fair allocation \cite{AIMD} is usually different from the desired bit-rate of each user application. 
In other words, one of the primary features in SDN to ensure the end-to-end bandwidth allocation may not sustain when end-to-end mouse flows are aggregated inside SDN.
However, many mouse flows such as Youtube streaming requires the sufficient bandwidth allocation (e.g., FULL HD).
To address the above issue, a possible approach is to facilitate flow rate control in every source client. However, this approach is not practically feasible because malicious users can easily overwhelm the whole network, and the network-based rate control is thereby necessary for SDN, as shown in the literature \cite{OpenFlowSwitch, SDTCP, FlowQoS, CoFlow}. In other words, it is desirable to provide a scalable network-based rate control mechanism for SDN mouse flows to support traffic engineering (TE) with aggregated flows.

\begin{figure}
    \centering
        \subfigure[]{%{0.45\linewidth}
            \includegraphics[width=4cm]{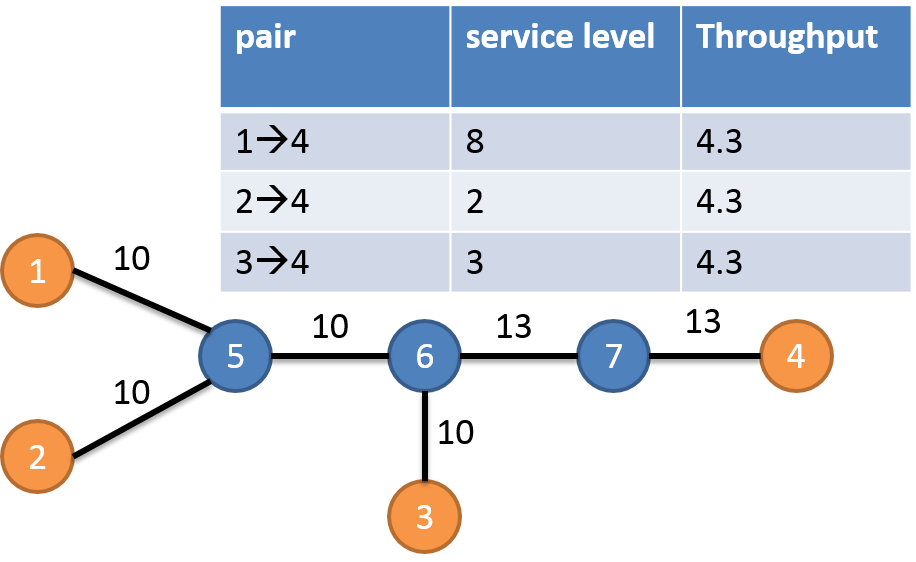}}
            %\label{fig:subim1}
        \hfill
        \subfigure[]{%{0.45\linewidth}
            \includegraphics[width=4cm]{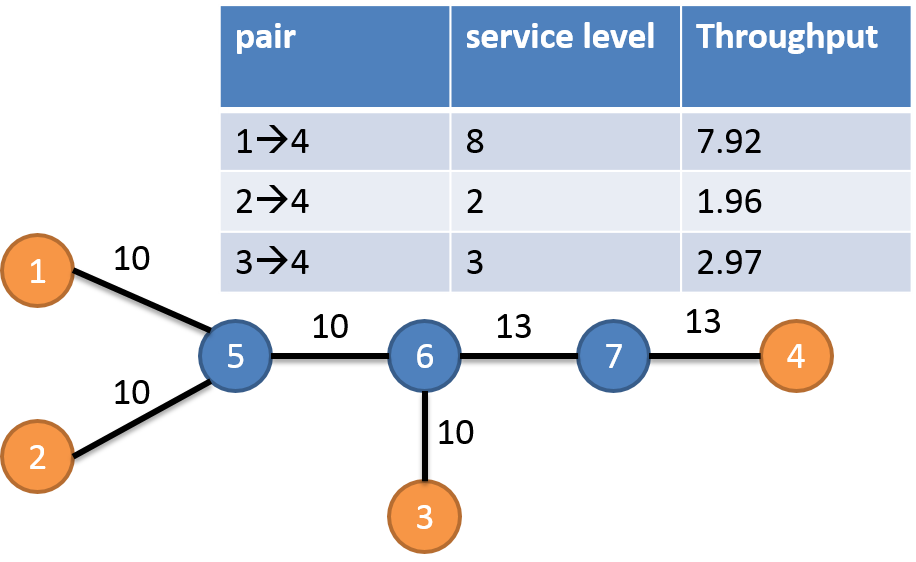}}
        \caption{\small{Uncontrolled and partially-controlled}}
        \label{fig: proof of concept}
\end{figure}

To address the above crucial need, this paper presents a new architecture, named Flexible Flow And Rate Management (F$^2$ARM). Given the routing of each mouse flow, our idea is to select only a few mouse flows to control the corresponding flow rates, whereas the rates of other flows are still controlled by TCP. In other words, only some mouse flows are selected by F$^2$ARM and controlled by SDN switches, and the other uncontrolled mouse flows are aggregated and share the aggregated bandwidth by TCP. F$^2$ARM in the SDN controller is designed to minimize the number of controlled flows in order to increase the scalability of SDN. Meanwhile, it guarantees that the bandwidth decided by TCP for every uncontrolled mouse flow exceeds its minimum user requirement. 
Fig. \ref{fig: proof of concept} presents the proof of concept in our small-scale environmental environment. Fig. \ref{fig: proof of concept}(a) shows that as all of the flows are uncontrolled (i.e., the bandwidth is allocated by the TCP), the allocated bandwidths of them are identical, 4.3, since the link $6\rightarrow 7$ is the bottleneck link. Fig. \ref{fig: proof of concept}(b) shows that as the flow $1\rightarrow 4$ is controlled while the other two flows are uncontrolled, by the experiment, all demands of the flows are almost satisfied.\footnote{Here we control the flow by a queue in OpenFlow.}
That is, partial controlling is a feasible way in SDN.
%Note that the traditional per-flow TE and per-flow admission control are special cases of F$^2$ARM, because every mouse flow is controlled in the above cases and thereby limiting the scalability of SDN, whereas only a small subset of flows needs to be controlled in F$^2$ARM. 
The remainder of this paper is structured as follows. Section \ref{sec: related work} summarizes the related work. Section \ref{sec: problem} provides the problem definition and hardness result. Section \ref{sec: algorithm} presents the proposed algorithm. Section \ref{sec: performance evaluation} shows the simulation results of our algorithm. Finally, we conclude the paper in section \ref{sec: conclusion}.

%\vspace{-0.2cm}
\section{Related Work}\label{sec: related work}
    \subsection{TE in SDN}
  
    Recently, flow aggregation has been explored for exploiting limited TCAM to achieve the scalable SDN. OFFICER \cite{OFFICER} reduced the TCAM consumption by routing many segments of flows into default paths between nodes. Bhatia et al. \cite{Segment01} properly determined the optimal routing parameters for each default path (i.e., segment) in both the offline and online cases, whereas Hao et al. \cite{Segment02} further investigated restoration optimization for network failure in the segment routing. Moreover, to efficiently exploit the limited TCAM, a novel rule multiplexing scheme \cite{RuleMultiplex} was proposed to minimize the rule space in TCAM. Meiners et al. \cite{bitweaving} proved that the non-prefix aggregation problem is NP-Hard and proposed to compress a given classifier into a non-prefix ternary classifier. Luo et al. \cite{aggregation} introduced the first online non-prefix aggregation scheme to shrink the flow-table size and support fast online updates. However, none of the above research focused on scalable rate control for aggregated flows in SDN.
    \subsection{TCP}
    To support scalable rate control in the Internet, many TCP variations \cite{Reno, NewReno, BIC, CUBIC, Illinois} have been proposed to achieve fair allocation of the network bandwidth by AIMD \cite{AIMD}. TCP-Reno \cite{Reno} first introduced fast recovery to increase the network connection utilization, while TCP-NewReno \cite{NewReno} further addressed multiple packet losses. BIC \cite{BIC} employed two window size control policies to alleviate unfairness of round-trip time (RTT).
    On the other hand, CUBIC \cite{CUBIC} simplified BIC by replacing the concave and convex window growth with a cubic function. TCP-Illinois \cite{Illinois} exploited the queueing delay to determine an increase factor and multiplicative decrease factor instantaneously during the congestion. Recently, fine-grained per-flow control with TCP in SDN switches has drawn increasing attention. FlowQoS \cite{FlowQoS} delegated application identification and QoS configuration from user clients to SDN controller for achieving per-flow QoS, while OMCoFlow \cite{CoFlow} minimized the average co-flow completion time in SDN. However, the above research targeted on per-flow rate control on individual flows with different objectives, instead of scalable rate control of the whole SDN with aggregated flows.
%\vspace{-0.2 cm}
\section{Problem and Hardness } \label{sec: problem}
%In this section, we first introduce the system model and formulate the SPFRCS problem with Mixed Integer Linear Programming (MILP).
%Afterward, we prove that 
%In addition, SPFRCS is NP-Hard and inapproximable within $(1-\epsilon)\ln \frac{|V|}{2}$.

%\subsection{Model and Formulation}
The network is modeled as a directed graph $G=\{V,E\}$, where $V$ and $E$ denote the sets of SDN switches and links, respectively. Each link $e\in E$ has an individual limited link capacity $c(e)\in\mathbb{R}^+$. Let $F$ denote the set of flows, where each flow $f\in F$ with a bandwidth demand $d(f)$ is assigned a routing path $p(f)$. %Each controlled flow is installed the flow state in at least one SDN switch in the path to control the flow rate. 
The state information of each controlled flow is maintained for rate control.
In contrast, the flow rates of uncontrolled flows are decided by TCP. 

In the SPFRCS problem, each flow can be either controlled or uncontrolled. Once it is controlled, the allocated bandwidth can be decided by the controller; otherwise, the allocated bandwidth is decided by TCP. Each uncontrolled flow has a bottleneck link that limits the allocated bandwidth of the flow. In addition, every link $e$ that has a uncontrolled flow will exhaust its bandwidth, and all of the flows whose bottleneck link is $e$ will equally share the bandwidth.
%\subsection{The Hardness}
%In this subsection, 
%we prove that
It is trivial that SPFRCS is NP-hard. It can be proved by a reduction from the set cover problem \cite{SetCoverInapproxRatio}. We have the following theorem.

%\begin{defi} \label{defi: set cover}
% Given a universe set $U$ of $n$ elements, a collection $S$ of $m$ sets whose union equals to the universe, the set cover problem is to identify the smallest sub-collection of $S$ that covers all elements of $U$.
%\end{defi}

\begin{theo} The SPFRCS problem is NP-hard.\label{theo: NP-hard}
\end{theo}
Since the set cover problem cannot be approximated within a factor of $(1-\epsilon)\ln n$, where $\epsilon>0$ and $n$ denotes the number of elements \cite{SetCoverInapproxRatio}, we have the following corollary.
\begin{coro}
The SPFRCS problem cannot be approximated within a factor of $(1-\epsilon)\ln \frac{|V|}{2}$, where $\epsilon > 0$ and $|V|$ denotes the number of switches in $G$.
\end{coro}
%\vspace{-0.2 cm}
\section{Algorithm}\label{sec: algorithm}
%In this section, 
%We first introduce the concept of algorithm JFSRD in Section \ref{subsec: algorithm concept} and then explain the details in Section \ref{subsec: algorithm details}. 
%Due to the space constraint, the pseudocode for JFSRD and the implementation details for deploying the rate control state in SDN are presented in \cite{TechnicalReport}.
    %\begin{figure}
    %\centering 
    %    \includegraphics[width=8cm]{images/TCPExample.pdf}
%        \label{fig:Example Case}
    %    \caption{\small{An example for TCP fairness}}
    %    \label{fig: TCP example}
    %\end{figure}
    \subsection{Algorithm Concept} \label{subsec: algorithm concept}
    To solve the problem, we propose algorithm Joint Flow Selection and Rate Determination (JFSRD) for SPFRCS to minimize the number of controlled flows in SDN. 
    %After controlled flow rates are extracted, TCP fair allocation is applied to assign the residual bandwidth to the uncontrolled flows. %, \textcolor{red}{according to the identities of uncontrolled flows in every links.}
    %Intuitively, 
    JFSRD first selects suitable controlled flows and then sets their rates to their minimum requirements. 
    After controlled flow rates are extracted, TCP fair allocation is applied to assign the residual bandwidth to the uncontrolled flows.
    Then, JFSRD further reduces the number of controlled flows by adjusting the controlled flow rates. % to limit the upper bound of  \textcolor{red}{the total uncontrolled flow rate in each link.}
    It is necessary to ensure that the allocated flow rate of each uncontrolled flow satisfies the minimum user requirement. 
    Therefore, JFSRD consists of two phases: 1) \textit{Flow Selection (FS)}, and 2) \textit{Rate Determination (RD)} 
    \textcolor{red}{
    %and one examined procedure:\textit{TCP procedure} 
    }
    
    %Intuitively,  \textcolor{red}{JFSRD first selects some controlled flows by iteratively examining the overlapping links among different sets of flows.} Then, JFSRD reduces the number of controlled flows by adjusting the controlled flow rates to limit the upper bound of  \textcolor{red}{the total uncontrolled flow rate in each link. It is necessary to ensure that the allocated flow rate of each uncontrolled flow satisfies the minimum user requirement.
     
    %More specifically, JFSRD consists of two phases: 1) \textit{Flow Selection (FS)}, and 2) \textit{Rate Determination (RD)}. In the following, we first describe several important guidelines for JFSRD. First, it is crucial to select the controlled flows by jointly considering the overlapping links of different flows. 
    %Note that each flow usually passes through multiple links to its destination, and each link of the path tends to be traversed by different sets of flows.
    %For each link, the flow rate of each controlled flow will affect the flow rates of the uncontrolled flows that regard the link as the bottleneck link. For each uncontrolled flow in the link, if its bottleneck link is another link, it will be assigned a smaller flow rate by TCP; otherwise, this link would be the bottleneck link. 
    \subsection{Algorithm Details}\label{subsec: algorithm details}
     To obtain a good solution for SPFRCS, we propose an ordering scheme to iteratively extract the controlled flows and assign their data rates by considering the correlation among different flows. JFSRD includes the following two phases.
    
    \noindent\textit{1) Flow Selection (FS)}
    
    JFSRD aims to find a \textit{configuration} (i.e., controlled or not) for all flows in the network. However, enumerating all possibilities (i.e., every flow can be controlled or not) is very computation intensive. Therefore, in order to reduce the computation complexity while ensuring the quality of solution, JFSRD first finds a set of candidate configurations instead of every possible configuration of the flows on each link, and then selects the most suitable configuration among the candidate set for each link. More specifically, the FS phase includes two stages, 1) \textit{configuration} and 2) \textit{selection}.% In the following, we first define the correlation of link $e$, $\gamma(e)$, which denotes the number of overlapping flows for the flows on $e$:
%    \begin{align}
%       &\gamma(e) = |\bigcup_{f\in F(e)}\{f'| p(f) \cap p(f') \neq \emptyset \wedge f'\in F\}|& \label{correlation}
%    \end{align} 

    In the \textit{configuration} stage, JFSRD initially assigns the data rate $r_f$ of each flow $f$ on $e$ as $d(f)$. The residual bandwidth of $e$, $\rho(e)$, is acquired accordingly, i.e., $\rho(e) = c(e) - \sum_{f:p(f) \ni e} r_f$. Afterward, in order to minimize the number of controlled flows in each configuration for each link $e$, JFSRD maximizes the number of flows with the same rates on $e$, so that many of them are potential to be uncontrolled flows via TCP. 
    \textcolor{red}{
    %The detail of this characteristic, we can see Fig. \ref{fig: TCP example} presents an example for TCP fairness allocation with 3 nodes, 3 edges and 5 flows, where the numbers beside each link is the link capacity. In the TCP, each flow respective increases their rate until an exhausted link appears and the flow goes through the exhausted link will stop increase their rate (accurately, the flow will oscillate between the stop rate). In this example, $e1$ will first be exhausted and the rate of $f1$ and $f2$ will stop at 2. Next, $e3$ will also be exhausted and the rate of $f3$, $f4$ and $f5$ will stop at 4. Based on the result of this example, we can observe that the flow with the same bottleneck link will have the same rate because of the fairness property of TCP.  
    } 
    Therefore, for link $e$, JFSRD iteratively picks a target rate in ascending order. For each target rate, JFSRD generates a candidate configuration for $e$ by distributing the residual bandwidth to the flows with a smaller rate, in order to maximize the number of flows with the rate identical to the target rate. To limit the number of configurations for each link $e$, JFSRD picks the target rates for each link $e$ from the rates of the flows on $e$. The number of configurations for each link $e$ is at most $|F(e)|$ since there are at most $|F(e)|$ flows on $e$.

    Afterward, in the \textit{selection} stage, an intuitive approach is to randomly pick a candidate configuration for each link. However, this approach tends to generate more controlled flows since it does not consider the common controlled flows among different links.
    Furthermore, the selected configuration of a more important link (i.e., the link visited by a larger number of flows and including more overlapping flows) may have more opportunities to significantly affect the total number of controlled flows.
    Therefore, JFSRD globally considers the importance of each link to decide its configuration. We define the \textit{correlation} of link $e$, $\gamma(e)$, which denotes the number of overlapping flows correlated to the flows on $e$. %, as follows:
%    \begin{align}
%        &\gamma(e) = |\bigcup_{f\in F(e)}\{f'| p(f) \cap p(f') \neq \emptyset \wedge f'\in F\}|& \label{correlation}
%    \end{align}
%    
    To address the above issue, therefore, JFSRD starts from the link $e$ with the highest correlation. %according to Eq. (\ref{correlation}).
    Subsequently, for a link, the candidate configuration with a larger proportion of controlled flows that are also controlled flows in the candidate configurations of other links has more opportunities to reduce the total controlled flows in the network. We call it the \textit{profit} of configuration $n$ on link $e$.
%    Intuitively, a candidate configuration with a higher profit implies  \textbf{ explain the reason, so why a larger profit implies fewer controlled flows?}. 
    As a result, JFSRD selects the configuration with the highest profit for $e$;
    that is, JFSRD prefers the configuration with a larger proportion of the controlled flows that are also controlled flows in the candidate configurations of other links.
    Afterward, JFSRD examines the next link with the highest correlation from all the links that have not been examined %. The FS phase repeats
    until the configurations of all links are selected. After the FS phase, each flow is decided to be controlled if and only if it is controlled in the selected configuration of any link. Note that the FS phase must satisfy the rate demands of all flows.
    Otherwise, assume that there is an unsatisfied uncontrolled flow $f$ whose bottleneck link is $e$. Then, it implies that the allocated rate of each uncontrolled flows on $e$ is at most the allocated rate of $f$, which is smaller than the target rate of the selected configuration of $e$.
    It contradicts since in such cases TCP does not exploit the available bandwidth of $e$ for each unsatisfied uncontrolled flow. 
    \textcolor{red}{
    %\begin{lemma}
    %    After FS phase, the $r_f$ of each flow in the network is greater or equal to the demand of $f$.
    %\end{lemma}
    %\begin{proof}
    %    If the flow $f$ status is controlled, the rate of $f$ is equal to $d(f)$. Otherwise, assume the flow $f$ is an unsatisfied uncontrolled flow and its bottleneck link is at $e$. Based on the TCP fairness property, we can know that if other uncontrolled flows go through the link $e$ their rate can only less or equal the rate of $f$. If the bottleneck link of other controlled flows is same as $e$ their rate is equal to $r_f$ otherwise their rate is less to $r_f$. In addition to, the configuration of $e$ which is selected after FS phase has a target rate which is greater or equal to the demand of each flow at $e$ for their uncontrolled flow and this target rate will not over-expend bandwidth of $e$. Thus, we can know that the $r_f$ and other uncontrolled flow is less than target rate and the bandwidth of $e$ will not be exhausted (i.e. $\sum_{f \in F^c_e(A_s(e))} d(f) + \sum_{f \in F(e) \setminus F^c_e(A_s(e))} r_f < c(e)$, where $A_s(e)$ denotes the selected configuration of each link $e \in E$ ). It contradicts since in such cases TCP does not exploit the available bandwidth of bottleneck link $e$ for each unsatisfied uncontrolled flow which bottleneck link is at $e$.      
    %\end{proof}
    }
    %TCP to exploit all available bandwidth for every uncontrolledflow
    
    %can be satisfied by the shared bandwidth on its bottleneck link.
    %stops when every flow is set as controlled or uncontrolled. 

%    \noindent\textit{2) TCP procedure}
    
    \noindent\textit{2) Rate Determination (RD)}
    
    In the RD phase, JFSRD further reduces the number of controlled flows that are selected in the FS phase. By observation, a flow $f$ that must be controlled implies that either 1) it will acquire the redundant bandwidth such that the other flows are unsatisfied or 2) it will not be satisfied, if it is uncontrolled. To solve condition 1 (or 2), a possible solution is to increase (or decrease) the rates of other controlled flows which visit a common link with $f$ such that the rate of $f$ can be bounded (or satisfied) without control. It is an implicit trade-off. Since the determined rate of each controlled flow in the FS phase cannot decrease anymore, JFSRD only increases the rates of controlled flows.
    More specifically, similarly, JFSRD starts from the link $e$ with the highest $\gamma(e)$ and then examines the \textit{temporarily controlled (TC)} flows of $e$, where \textit{TC} means that the flows are uncontrolled in the selected configuration on $e$, but they are controlled in the selected configurations on the other links. 
    These \textit{TC} flow rates can be limited by increasing the other controlled-flow rates on $e$. Therefore, it is envisaged that their rates can be assigned properly by TCP even they are not controlled. 
    To fairly allocate the bandwidth, JFSRD proportionally allocates the residual bandwidth of $e$ to the controlled flows according to their priority that can be specified by the network owner or applications. A flow with a higher priority will acquire more residual bandwidth. 
    Note that during the allocation, if there is a link to be exhausted on the path of some controlled flow $f$, the residual bandwidth of $e$ is then proportionally assigned to the other controlled flows (i.e., not $f$) on $e$ according to the priority.

    The above allocation process repeats until the residual bandwidth of $e$ is exhausted, or no controlled flow on $e$ can increase its rate. %After the residual bandwidth is exhausted, the \textit{TC} flow on $e$ becomes uncontrolled once its demand rate can be satisfied by TCP. On the other hand, 
    %If the residual bandwidth cannot be completely allocated, 
    Then, JFSRD examines whether the allocation can satisfy all the flow demands if the \textit{TC} flows on $e$ are set as uncontrolled flows \textcolor{red}{
    %by the \textit{TCP procedure}
    }. If all the flows can be satisfied, the \textit{TC} flows on $e$ are assigned as uncontrolled flows; otherwise, they become controlled flows. Afterward, JFSRD updates the status and examines the next link with the highest correlation among all the links that have not been examined. The RD phase stops when all the \textit{TC} flows on each link are examined.

%\vspace{-0.2 cm}
%\input{Discussion}
\section {Performance Evaluation}\label{sec: performance evaluation}
    \subsection{Simulation Setup}
    \begin{figure}
    \centering
        \subfigure[]{%{0.45\linewidth}
            \includegraphics[height=3cm]{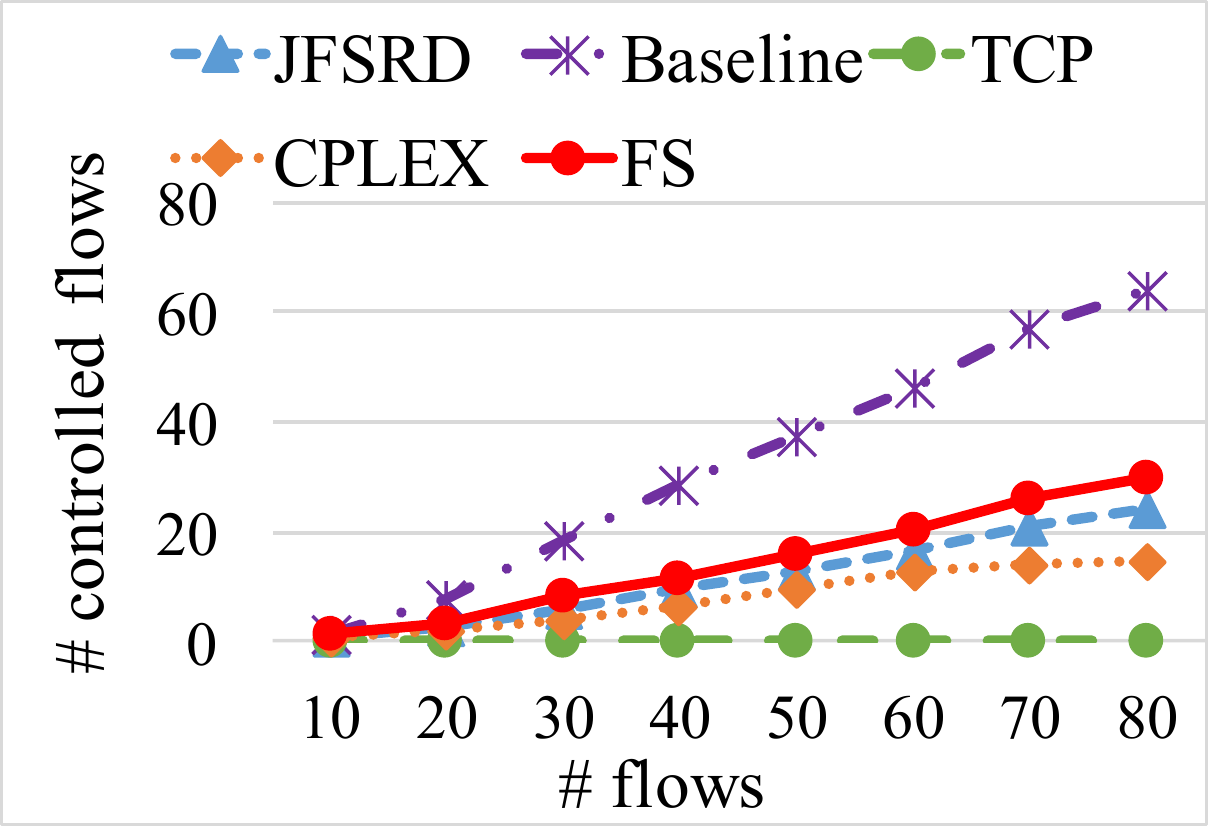}}
            %\label{fig:subim1}
        \hfill
        \subfigure[]{%{0.45\linewidth}
            \includegraphics[height=3cm]{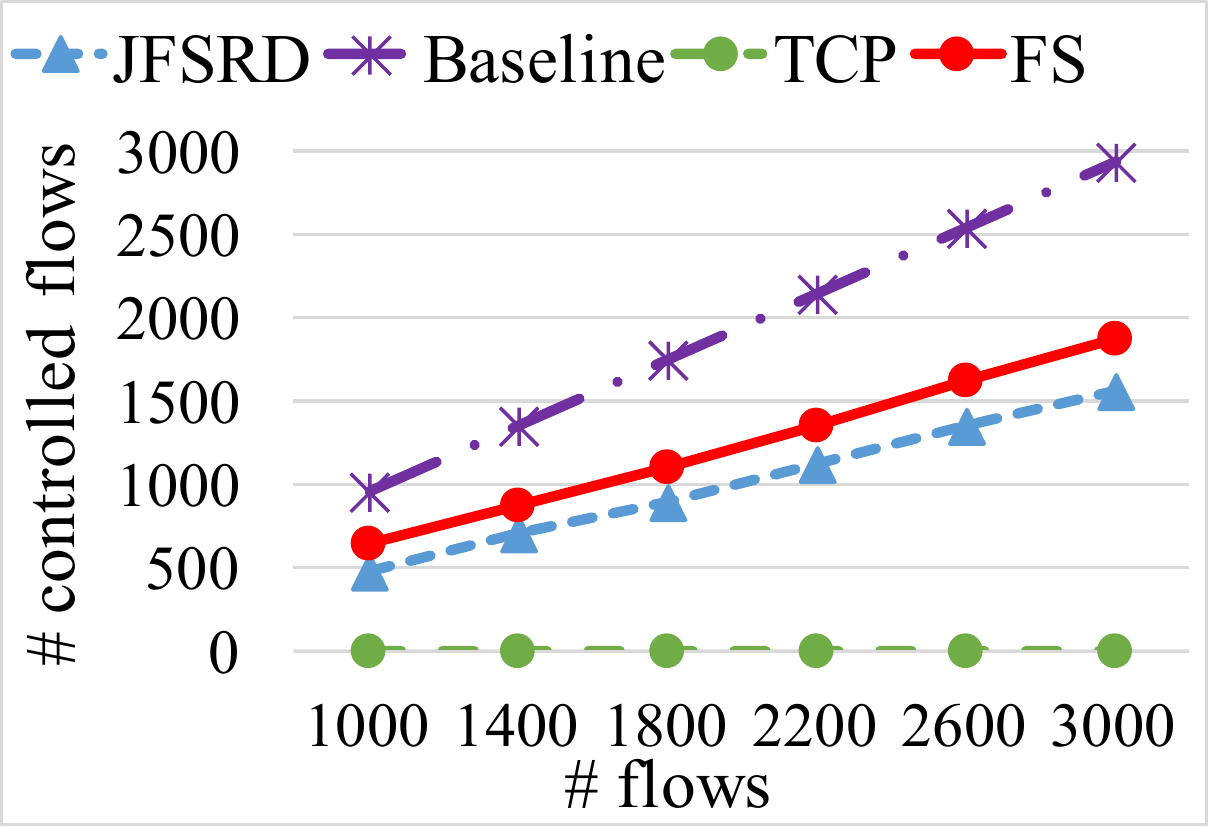}}
            %\label{fig:subim1}
%        \vspace{-0.3cm}
        \subfigure[]{%{0.45\linewidth}
            \includegraphics[height=3cm]{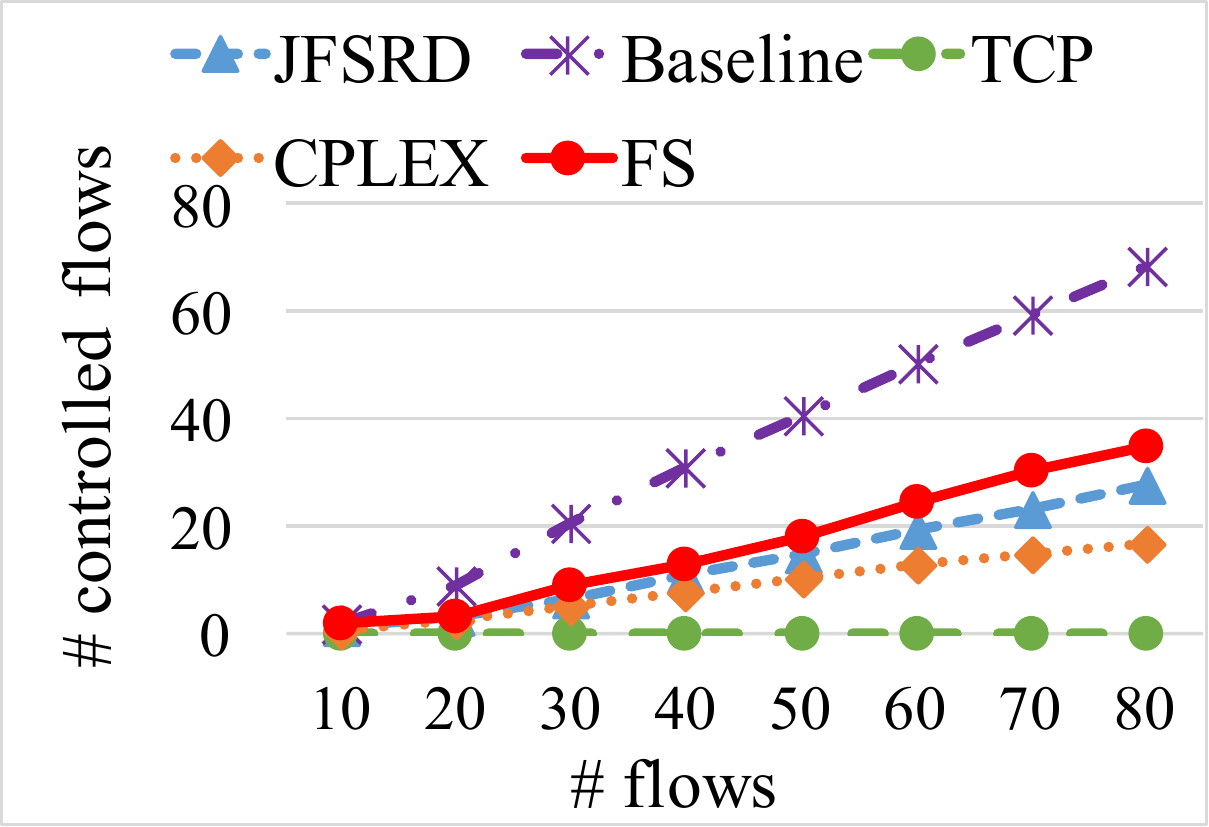}}
            %\label{fig:subim1}
        \hfill
        \subfigure[]{%{0.45\linewidth}
            \includegraphics[height=3cm]{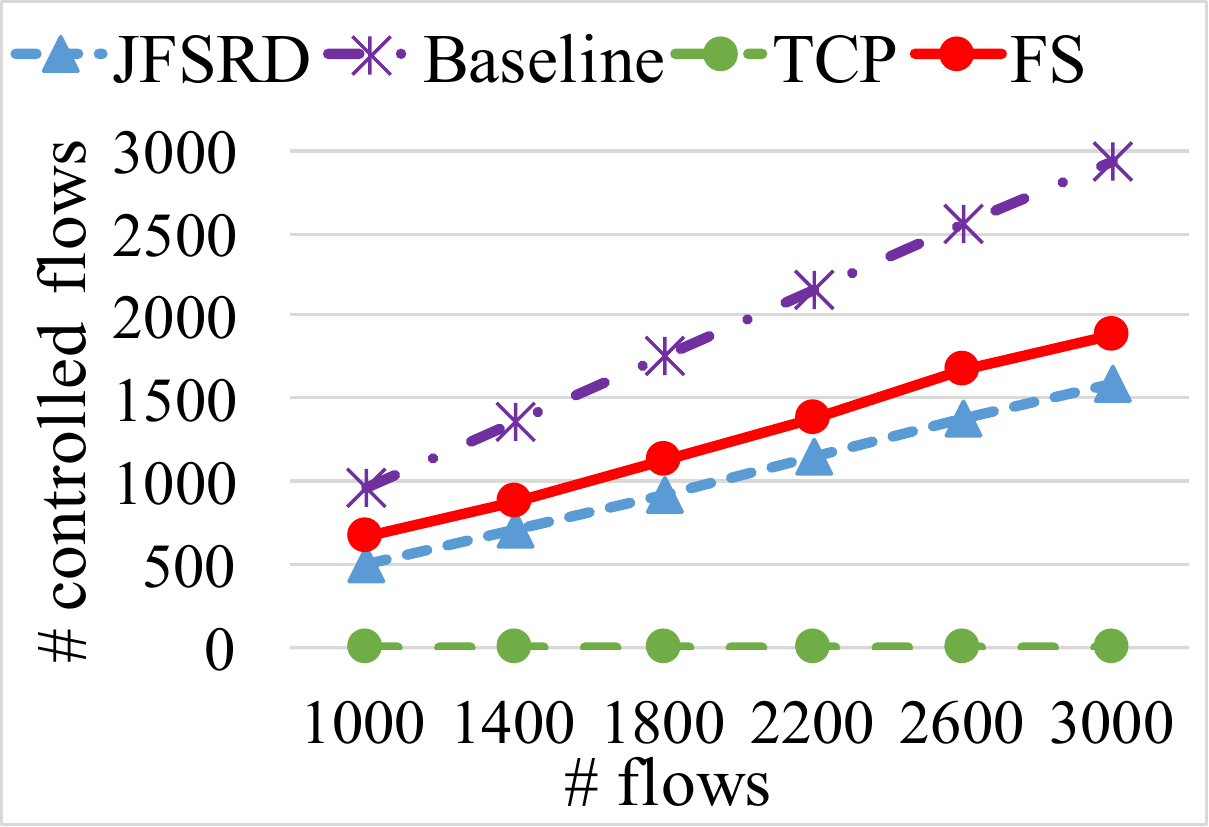}}
            %\label{fig:subim1}
        %\vspace{-0.5cm}
        \caption{\small{The performance of JFSRD with varying number of flows in (a) Claranet and (b) Columbus with shortest path, and (c) Claranet and (d) Columbus with OFFICER}}
    \label{fig: simulation 1}
    \end{figure}
    We conduct extensive simulations to evaluate the performance of JFSRD in two real networks: Claranet and Columbus \cite{TopologyZoo}. Claranet includes 15 nodes and 18 links, while Columbus has 70 nodes and 85 links. The number of flows in Claranet ranges from 30 to 80, whereas there are more than 1000 flows in Columbus. Two routing policies in SDN are tested in the simulation: shortest-path routing\cite{ShrotestPath} and OFFICER\cite{OFFICER} that supports SDN TE with aggregated flows. Since this paper targets on flow control, the link capacity is set to 25$\%$ more than the total flow demands in each link. The flow demands are generated by traffic matrices according to\cite{hybrid03}, and the source and destination of each flow are randomly chosen. 
    
    In the simulation, since there is no related work exploring scalable rate control for SDN TE with aggregated flows, we compare JFSRD with three algorithms: 1) traditional TCP, where the link bandwidth are shared by all flows, 2) the baseline algorithm that iteratively selects the flow with the smallest ID as a controlled flow until the remaining unselected flows can be satisfied by TCP, and 3) IBM CPLEX %12.6 \cite{CPLEX} 
    to find the optimal solutions of SPFRCS problem with the MILP formulation in Section \ref{sec: problem}. %III.
    %{\color{blue}
    Note that the running time of each instance by CPLEX is over ten minutes even if there are more than 80 source-destination pairs, and thus we do not compare JFSRD with the optimal solution in the large case -- Columbus.
    %}
    To evaluate JFSRD, we vary the following parameters: 1) number of flows and 2) load of networks. When the number of flows varies, the average load of the network is fixed at 75$\%$. When the load of networks varies, the number of flows is fixed at 60 and 1500 for Claranet and Columbus, respectively. We measure the following two performance metrics: 1) number of controlled flows and 2) number of flows not meeting the user requirements. Each  result is averaged over 100 samples.
\begin{figure}
    \centering
        \subfigure[]{%{0.45\linewidth}
            \includegraphics[height=3cm]{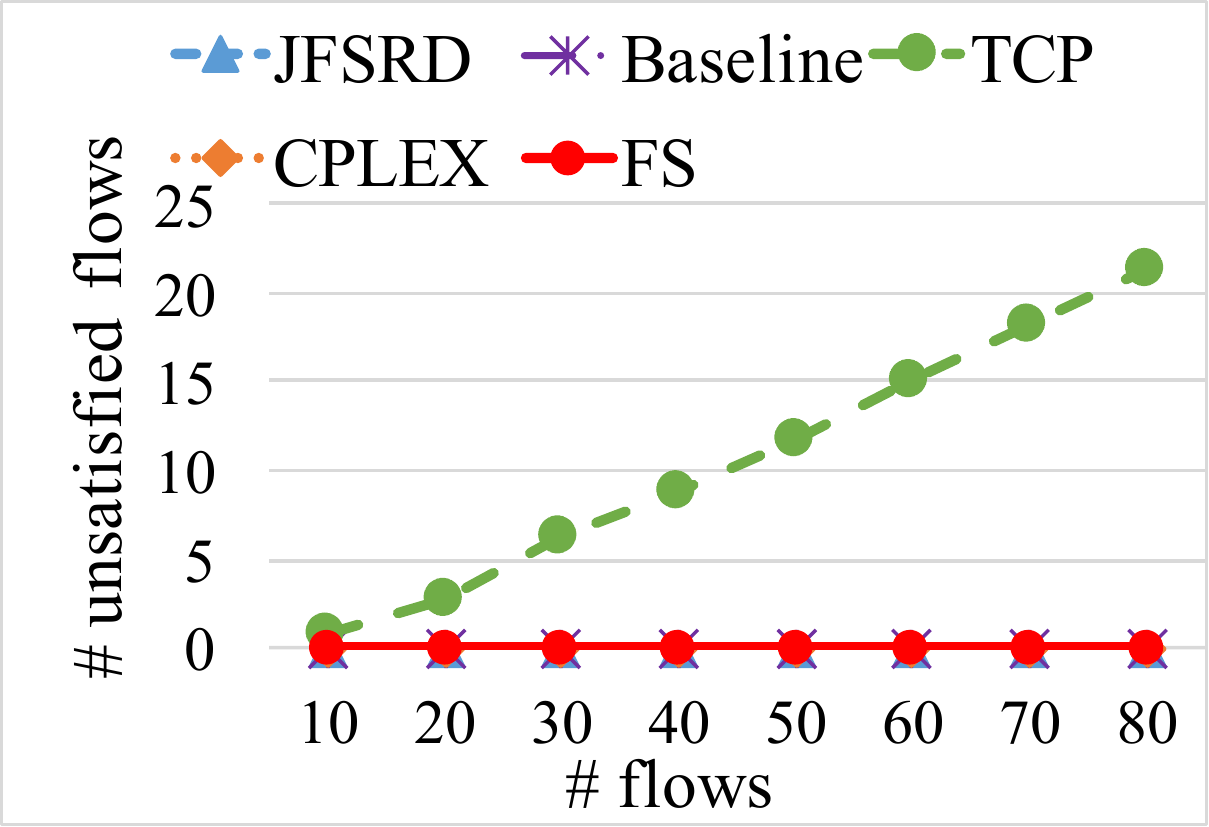}}
            %\label{fig:subim1}
        \hfill
        \subfigure[]{%{0.45\linewidth}
            \includegraphics[height=3cm]{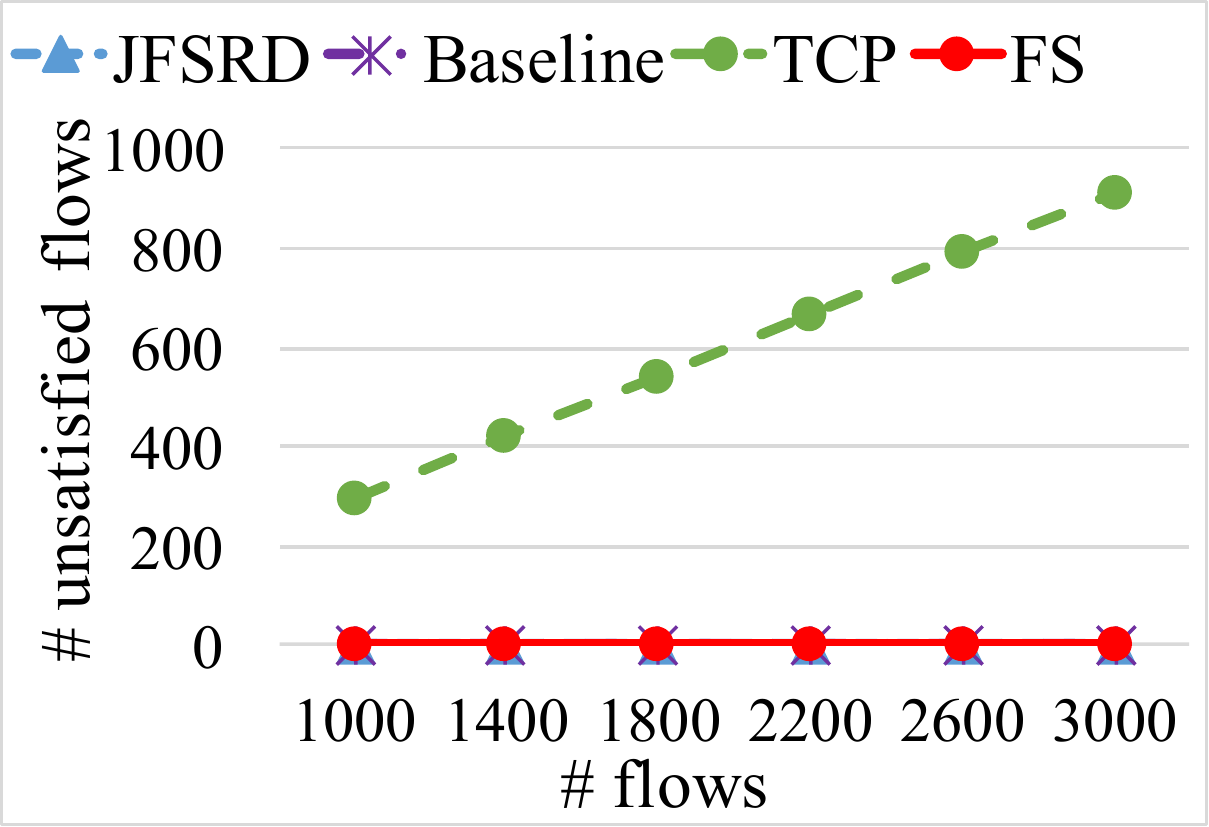}}
%            \label{fig:subim1}
        \subfigure[]{%{0.45\linewidth}
            \includegraphics[height=3cm]{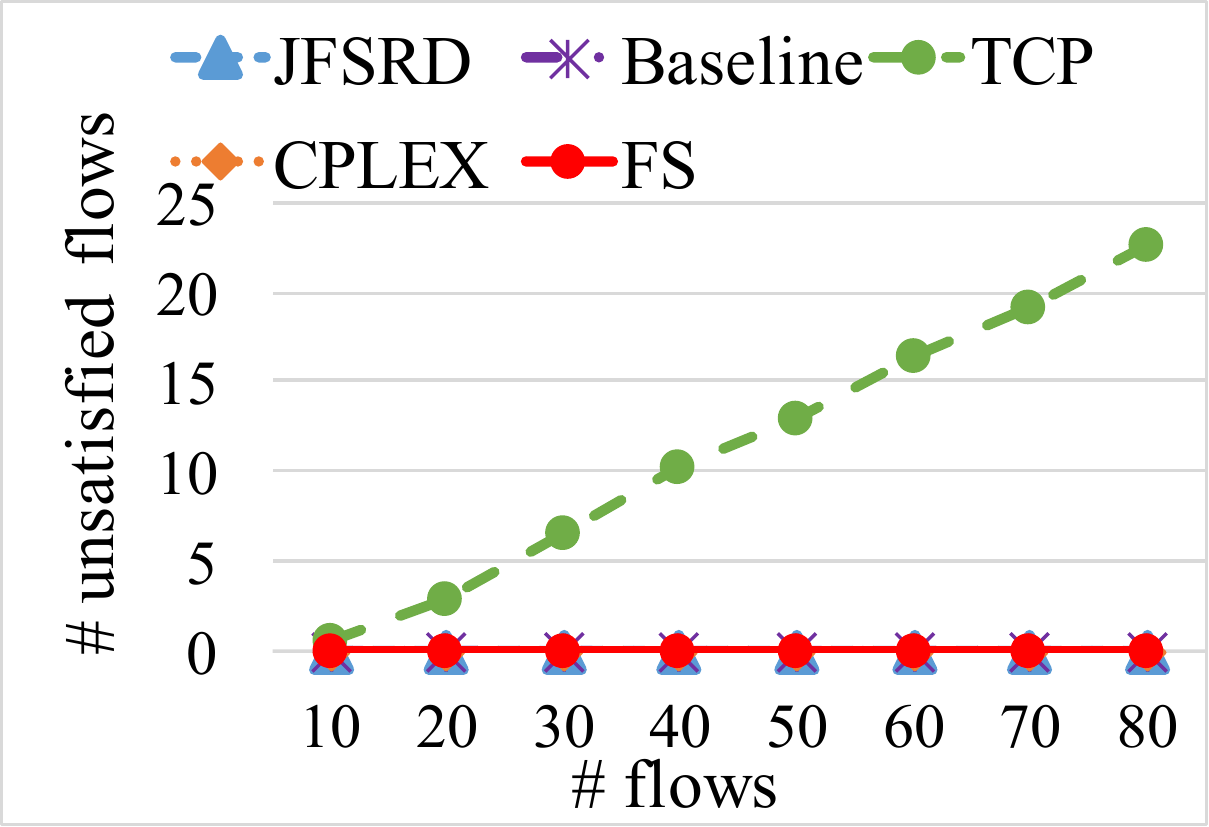}}
            %\label{fig:subim1}
        \hfill
        \subfigure[]{%{0.45\linewidth}
            \includegraphics[height=3cm]{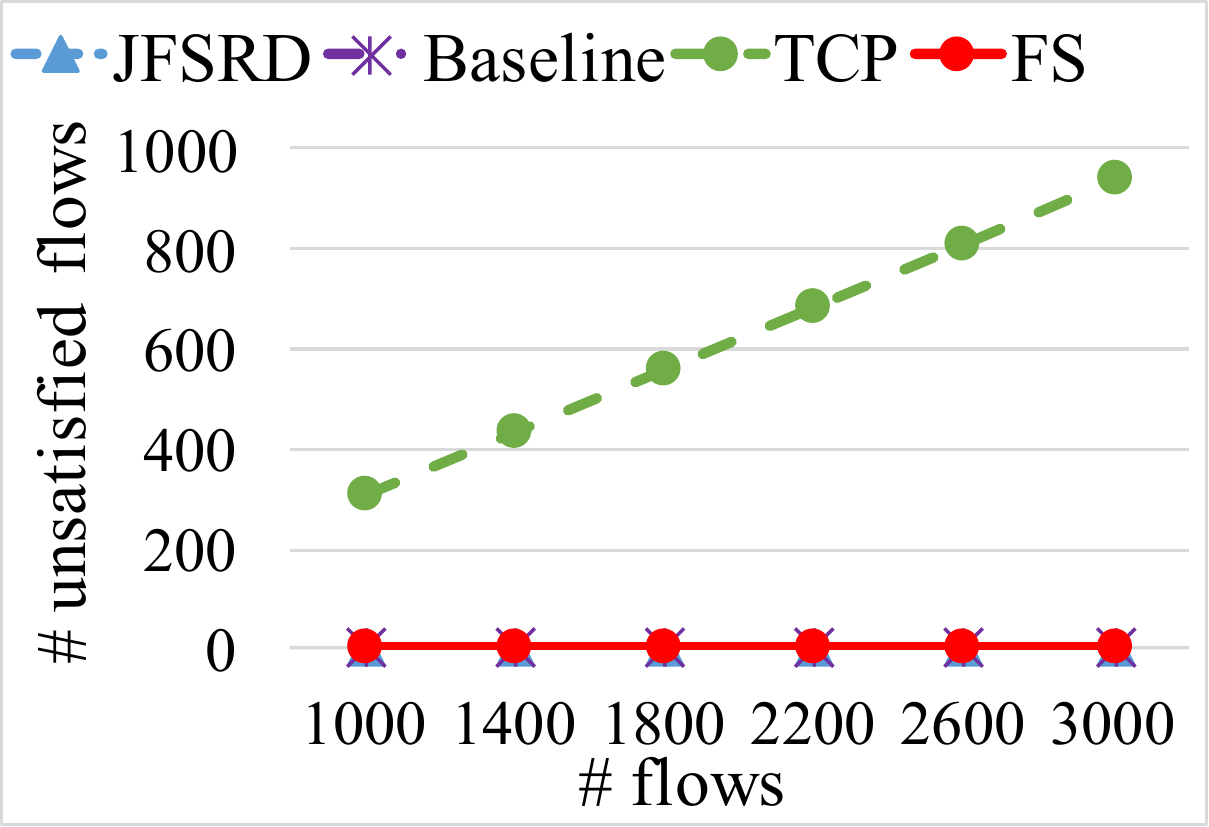}}
%            \label{fig:subim1} 
        %\vspace{-0.5cm}
        \caption{\small{The number of unsatisfied flows in (a) Claranet and (b) Columbus with shortest path, and (c) Claranet and (d) Columbus with OFFICER}}
        \label{fig: simulation 2}
\end{figure}
    \subsection{Simulation Result}
    Figs. \ref{fig: simulation 1} and \ref{fig: simulation 2} manifest that JFSRD greatly outperforms the baseline approach over 50 $\%$ on the number of controlled flows in almost all cases, and all user requirements can be satisfied. Moreover, the solution of JFSRD is very close to the optimal solution in Claranet because JFSRD selects the controlled flows and adjusts the flow rates according to the correlation of overlapping links among different sets of flows, instead of simply iteratively selecting the controlled flows based on their ID and setting the rate as their demands. It is worth noting that the FS phase in JFSRD has already obtained a fairly good solution. In addition, for TCP nearly 25 $\%$ of flows cannot meet the corresponding rate requirements. 
    
    In the following, we first investigate the number of flows versus the number of controlled flows with two routing schemes in SDN. Fig. \ref{fig: simulation 1} first shows that JFSRD significantly outperforms the baseline approach for both routing policies in both networks and generates the solution very close to the optimal solution in Claranet network.
    This is because JFSRD attempts to select the minimal number of controlled flows for each link, and it also carefully examines the number of links that can benefit from the selection. 
    In other words, JFSRD selects the flows that are usually the common controlled flows of several configurations in different links.
    On the other hand, the baseline approach selects a huge number of unnecessarily controlled flows due to the naive ID-based selection, and it usually misses some flows that are highly correlated to the other flows to limit their rates on some links.
    Moreover, Fig. \ref{fig: simulation 1} and Fig. \ref{fig: simulation 2} demonstrate that although TCP automatically control the flow rates, 25\% of the users are not satisfied since it fairly allocates bandwidth to the flows without considering their rate demands.
    %The TCP in Fig. 3 presents that they don't result in a burden on the controller but in Fig. 4 clearly indicate they make nearly 25 percentage of flows can't reach their Qos requirement. 
    Also,
    Fig. \ref{fig: simulation 1} (a) and (c) (or (b) and (d)) manifest that the number of controlled flows with OFFICER TE is slightly larger than that with shortest-path routing because OFFICER tends to route the traffic along the default path toward the controller before it reaches the selected branch switch. Therefore, it produces more highly-utilized links in the network. 
\section {Conclusion}
\label{sec: conclusion}
To the best of our knowledge, this paper is the first attempt to minimize the number of controlled flows by assigning the rate of subset of flows so that the rates for uncontrolled flows by TCP can exceed the corresponding user requirements. We first introduce F$^2$ARM for the SDN controller to extract a subset of controlled flows in order to increase the scalability of SDN, while the remaining uncontrolled flows are managed by TCP. However, the selection of controlled flows is challenging since an uncontrolled flow may meet different competing flows in different links to share the bandwidth. Therefore, we formulate a new optimization problem, SPFRCS, and prove that the hardness of the problem. To solve the problem, we devise an effective algorithm JFSRD to iteratively select the controlled flows and determine their rates by carefully examining the overlapping links among different sets of flows. Simulation results demonstrate that the number of controlled flows can be effectively reduced by $50\%$ in most cases, and the performance of JFSRD is very close to the optimal solution in real networks. 
%\vspace{-0.2 cm}
%\input{reference}

\ifCLASSOPTIONcaptionsoff
  \newpage
\fi

\bibliographystyle{IEEEtran}
\bibliography{reference}

\begin{thebibliography}{10}
\providecommand{\url}[1]{#1}
\csname url@samestyle\endcsname
\providecommand{\newblock}{\relax}
\providecommand{\bibinfo}[2]{#2}
\providecommand{\BIBentrySTDinterwordspacing}{\spaceskip=0pt\relax}
\providecommand{\BIBentryALTinterwordstretchfactor}{4}
\providecommand{\BIBentryALTinterwordspacing}{\spaceskip=\fontdimen2\font plus
\BIBentryALTinterwordstretchfactor\fontdimen3\font minus
  \fontdimen4\font\relax}
\providecommand{\BIBforeignlanguage}[2]{{%
\expandafter\ifx\csname l@#1\endcsname\relax
\typeout{** WARNING: IEEEtran.bst: No hyphenation pattern has been}%
\typeout{** loaded for the language `#1'. Using the pattern for}%
\typeout{** the default language instead.}%
\else
\language=\csname l@#1\endcsname
\fi
#2}}
\providecommand{\BIBdecl}{\relax}
\BIBdecl

\bibitem{OpenFlow}
N.~McKeown, T.~Anderson, H.~Balakrishnan, G.~Parulkar, L.~Peterson, J.~Rexford,
  S.~Shenker, and J.~Turner, ``Openflow: Enabling innovation in campus
  networks,'' \emph{ACM SIGCOMM Comput. Commun. Rev.}, vol.~38, pp. 69--74,
  2008.

\bibitem{AIMD}
F.~Baccelli and D.~Hong, ``Aimd, fairness and fractal scaling of tcp traffic,''
  in \emph{IEEE INFOCOM}, 2002.

\bibitem{OpenFlowSwitch}
\BIBentryALTinterwordspacing
{OpenFlow} switch specification. [Online]. Available:
  \url{https://www.opennetworking.org/technical-communities/areas/specification}
\BIBentrySTDinterwordspacing

\bibitem{SDTCP}
Y.~Lu and S.~Zhu, ``Sdn-based tcp congestion control in data center networks,''
  in \emph{IEEE IPCCC}, 2015.

\bibitem{FlowQoS}
M.~S. Seddiki, M.~Shahbaz, S.~Donovan, S.~Grover, M.~Park, N.~Feamster, and
  Y.-Q. Song, ``{FlowQoS}: {QoS} for the rest of us,'' in \emph{ACM HotSDN},
  2014.

\bibitem{CoFlow}
Y.~Li, S.~H.-C. Jiang, H.~Tan, C.~Zhang, G.~Chen, J.~Zhou, and F.~C.~M. Lau,
  ``Efficient online coflow routing and scheduling,'' in \emph{ACM MobiHoc},
  2016.

\bibitem{OFFICER}
X.~N. Nguyen, D.~Saucez, C.~Barakat, and T.~Turletti, ``Officer: A general
  optimization framework for openflow rule allocation and endpoint policy
  enforcement,'' in \emph{IEEE INFOCOM}, 2015.

\bibitem{Segment01}
R.~Bhatia, F.~Hao, M.~Kodialam, and T.~V. Lakshman, ``Optimized network traffic
  engineering using segment routing,'' in \emph{IEEE INFOCOM}, 2015.

\bibitem{Segment02}
F.~Hao, M.~Kodialam, and T.~V. Lakshman, ``Optimizing restoration with segment
  routing,'' in \emph{IEEE INFOCOM}, 2016.

\bibitem{RuleMultiplex}
H.~Huang, S.~Guo, P.~Li, B.~Ye, and I.~Stojmenovic, ``Joint optimization of
  rule placement and traffic engineering for qos provisioning in software
  defined network,'' \emph{IEEE Trans. Comput.}, vol.~64, pp. 3488--3499, 2015.

\bibitem{bitweaving}
C.~R. Meiners, A.~X. Liu, and E.~Torng, ``Bit weaving: A non-prefix approach to
  compressing packet classifiers in tcams,'' \emph{IEEE/ACM Trans. Netw.},
  vol.~20, pp. 488--500, 2012.

\bibitem{aggregation}
S.~Luo, H.~Yu, and L.~M. Li, ``Fast incremental flow table aggregation in
  sdn,'' in \emph{ICCCN}, 2014.

\bibitem{Reno}
V.~Jacobson, ``Berkeley tcp evolution from 4.3-tahoe to 4.3-reno,'' in
  \emph{Proc. IETF}, 1990.

\bibitem{NewReno}
S.~Floyd and T.~Henderson, ``The newreno modification to tcp's fast recovery
  algorithm,'' RFC Editor, RFC 2582, 1999.

\bibitem{BIC}
L.~Xu, K.~Harfoush, and I.~Rhee, ``Binary increase congestion control (bic) for
  fast long-distance networks,'' in \emph{IEEE INFOCOM}, 2004.

\bibitem{CUBIC}
S.~Ha, I.~Rhee, and L.~Xu, ``Cubic: A new tcp-friendly high-speed tcp
  variant,'' \emph{ACM SIGOPS Oper. Syst. Rev.}, vol.~42, pp. 64--74, 2008.

\bibitem{Illinois}
S.~Liu, T.~Başar, and R.~Srikant, ``Tcp-illinois: A loss- and delay-based
  congestion control algorithm for high-speed networks,'' \emph{Elsevier
  Performance Evaluation}, vol.~65, pp. 417 -- 440, 2008.

\bibitem{SetCoverInapproxRatio}
I.~Dinur and D.~Steurer, ``Analytical approach to parallel repetition,'' in
  \emph{ACM STOC}, 2014.

\bibitem{TopologyZoo}
\BIBentryALTinterwordspacing
The internet topology zoo. [Online]. Available: \url{http: //www.topology-
  zoo.org/dataset.html}
\BIBentrySTDinterwordspacing

\bibitem{ShrotestPath}
M.~Rifai, N.~Huin, C.~Caillouet, F.~Giroire, D.~Lopez-Pacheco, J.~Moulierac,
  and G.~Urvoy-Keller, ``Too many sdn rules? compress them with minnie,'' in
  \emph{IEEE GLOBECOM}, 2015.

\bibitem{hybrid03}
Y.~Hu, W.~Wang, X.~Gong, X.~Que, Y.~Ma, and S.~Cheng, ``Maximizing network
  utilization in hybrid software-defined networks,'' in \emph{IEEE GLOBECOM},
  2015.

\end{thebibliography}

%\input{Algorithm}

%\begin{IEEEbiography}[{\includegraphics[width=1in,height=1.25in,clip,keepaspectratio]{picture}}]{John Doe}
%\blindtext
%\end{IEEEbiography}
\end{document}